\documentclass[12pt,preprint]{aastex}

\usepackage{epsf}\usepackage{psfig}\usepackage{lscape}

\shorttitle{Effective gravity of halo stars}
\shortauthors{Bessell}

\begin{document}


\title{Measuring the Balmer jump and the effective gravity in FGK stars}


\author{Michael S. Bessell}
\affil{RSAA, The Australian National University,
    Mt Stromlo, Cotter Rd, Weston, ACT 2611, Australia}

\begin{abstract}
It is difficult to accurately measure the effective gravity (log g) in late-type
stars using broadband (eg. UBV or SDSS) or intermediate-band (uvby)
photometric systems, especially when the stars can cover a range of 
metallicities and reddenings. However, simple spectroscopic observational and 
data reduction techniques can yield accurate values for log g through comparison 
of the Balmer jumps of low-resolution spectra with recent grids of synthetic 
flux spectra. 

\end{abstract}




\section{Introduction}
The abundance analyses of the oldest stars in the galaxy are of great importance to
understand which elements were created during the life and death of the
first generation of stars and how the fraction of heavy elements built up through
the evolution of subsequent generations of stars. The effective temperature
is the most important stellar atmosphere parameter to assign in the determination
of abundances, but for those elements only represented by lines of the neutral 
species, it is necessary to establish the electron pressure or the effective
gravity of the star. This is because hotter than about 4500K most of the 
heavy elements are once ionised and significant corrections are required 
to determine the total element abundances from the neutral species. 
Theoretical isochrones for a range of metallicities and ages are 
readily available \citep{piet2006, vdb2000, sal2000} and the normal log 
($L/L_{\sun}$) versus log Te relations can be readily transformed to log g 
versus Te (log g = 4log $T_{e}$ - log($L/L_{\sun}$) - 10.608). 

\begin{figure}[h]
\centerline{\hbox{\psfig{figure=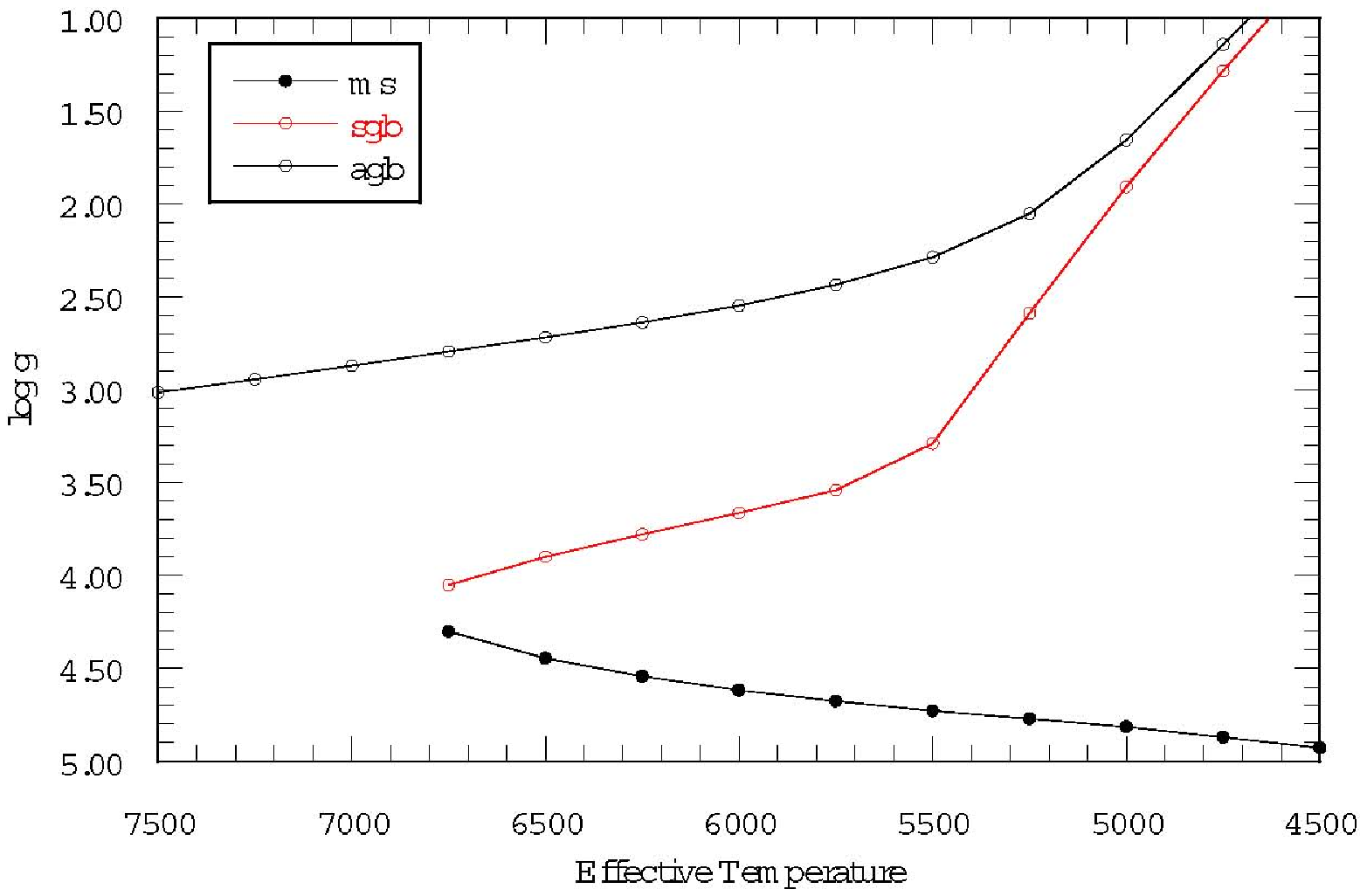,angle=0,clip=,width=13cm}}}
\caption[]{Isochrone from \citet{piet2006} for 12 Gyrs.}
\label{fig:f1}
\end{figure}

Fig.~\ref{fig:f1} shows a typical halo isochone for age 12 Gyr. 
For old stars (age $>$ 10 Gyr),
the shape of the isochrone changes little as the main-sequence turn-off 
(TO) moves to cooler temperatures with older age. As can be seen from 
the figure, for a given temperature, there are normally only three different 
possible values for log g for a halo star, depending on whether the star is 
on the main-sequence (MS), on the subgiant/giant branch (SGB) or on the 
horizontal branch (HB) or asymptotic giant branch (AGB). 
Cooler than about 5500K it is generally quite easy to choose between a MS 
(log g = 5), a GB (log g = 3.5) and a HB star (log g = 2.5) but near the MS 
turnoff the gravity differences are quite small and more difficult to discern. 

The Balmer discontinuity is an obvious feature to measure in the spectra 
of late-type halo stars in order to determine log g. Blueward of 3636\AA\/ the 
continuous opacity is mainly H-, whereas redward it is H- plus HI. As log
g (and log P$_{e}$) decreases, the H- opacity decreases and the size of
the Balmer discontinuity increases. But as the HI opacity decreases steeply 
with decreasing temperature, the Balmer jump becomes less sensitive to 
changes in log g.

Although low dispersion spectra are regularly taken of halo stars in order
to measure the HK lines of CaII and the H$\delta$ line for metallicity and
temperature estimates, little or no effort is put into measuring the Balmer
discontinuity. I think that there are three main reasons for this. Firstly, 
as the Balmer discontinuity is near the confluence of the hydrogen lines the
continuum is changing rapidly and it seems impossible to uniquivocally 
extrapolate a curved line that fits the continuum from the 4100\AA\/ to 
4800\AA\/ region where the continuum is well defined. Secondly, the 
sensitivity of most spectrographs and CCDs decreases rapidly below 
3900\AA\/ leading many people to think there is little point in observing 
below 3600\AA\/. And finally, many think that it may be possible to 
overcome these difficulties only if a full relative-absolute 
spectrophotometric calibration is undertaken and this seems like 
too much effort for uncertain return. 

This paper aims to show how, with a minimal amount of extra effort at both the 
telescope and with the data reduction, the Balmer jump can be readily 
measured and compared with similar data from the detailed grid of synthetic 
spectra provided by \citet{mun2005}.        
 
\section{Techniques}
\subsection{Atmospheric dispersion}
When doing ground-based spectroscopy in the UV/blue it is highly desirable 
that the effect of atmospheric dispersion is minimised either by using an 
atmospheric dispersion corrector \citep{fil1982} or by rotating the 
spectrograph slit so that it is parallel to the atmospheric dispersion. 
Otherwise, at large zenith distances, if the guider is centering the green or 
red image, a significant proportion of the UV/blue light will miss the slit. 
But, as the Balmer jump is measured at one wavelength, it is not 
absolutely essential to measure the same proportion of light at all wavelengths.
However, although a reasonable discontinuity may be measured under these 
circumstances, any possibility of measuring a temperature sensitive slope or 
color such as B-V is lost. If an atmospheric dispersion corrector is
unavailable, spectra should be taken with the slit tracking the parallactic 
angle or the star should be acquired in that mode and then observed with 
the position angle fixed to use an offset guide star in an alt-az telescope. 

\subsection{Division by black-body spectrum}
\citet{bes99} discusses the advantages of observing smooth spectrum stars
as part of any observing program. After the raw spectra have been extracted, 
the first step is to divide all spectra by the normalised spectrum of a star
with a near-black-body energy distribution. 

There are several near-continuous bright white dwarfs well suited for such 
division. These are \objectname{EG131} and \objectname{L745-46a} which are 
accessible in opposite seasons together
with \objectname{vMa2} and two carbon-rich white dwarfs \objectname{LHS 43}
\citep{bes99} and \objectname{LHS 4043} \citep{duf05}. It
is necessary to remove the weak C$_{2}$ bands from the carbon WDs
and the CaII HK and MgI lines from L745-46a and vMa2, or fit a line through the 
continuum before division. 
\citet{green90} and \citet{berg01} discuss the spectra of these stars while 
\citet{koes00} and \citet{duf05} discuss the theoretical fluxes of some of
the stars.

There are two main reasons for dividing by a normalised near-BB spectrum.
Firstly, in the red it removes most of the atmospheric (telluric) 
absorption lines and secondly, for both blue and red spectra it removes most
of the instrumental response due to vignetting, grating blaze, filters, mirrors 
and CCD response. The net effect of this division is to provide spectra 
in which the continuum level changes slowly and smoothly with wavelength that 
in turn means the corrections (with wavelength) to place the fluxes on
a relative absolute flux scale can be well fitted with a smooth low order 
polynomial or spline. 

\subsection{Linearisation of the continuum}
An unanticipated dividend of the division by a warm near-BB spectrum 
was the fact that the continuum levels of FGK stars above and below the 
Balmer jump were transformed from curves into almost straight lines. This
enables the continua to be extrapolated with confidence to 3636\AA\/ and the 
Balmer jump measured accurately and consistently.

Division by EG131, black-body temperature about 11800K, 
straightens the continuum in stars with temperatures from 7500K to 5000K.  
Division by the cooler L745-46a ($\approx$ 8600K) leaves some residual 
curvature for spectra hotter than 6500K but is good for cooler stars.
The two carbon-rich white dwarfs LHS43 and LHS4043 have intermediate 
black-body temperatures to EG131 and L745-46a; vMa2 is cooler.   

\begin{figure}[h]
\centerline{\hbox{\psfig{figure=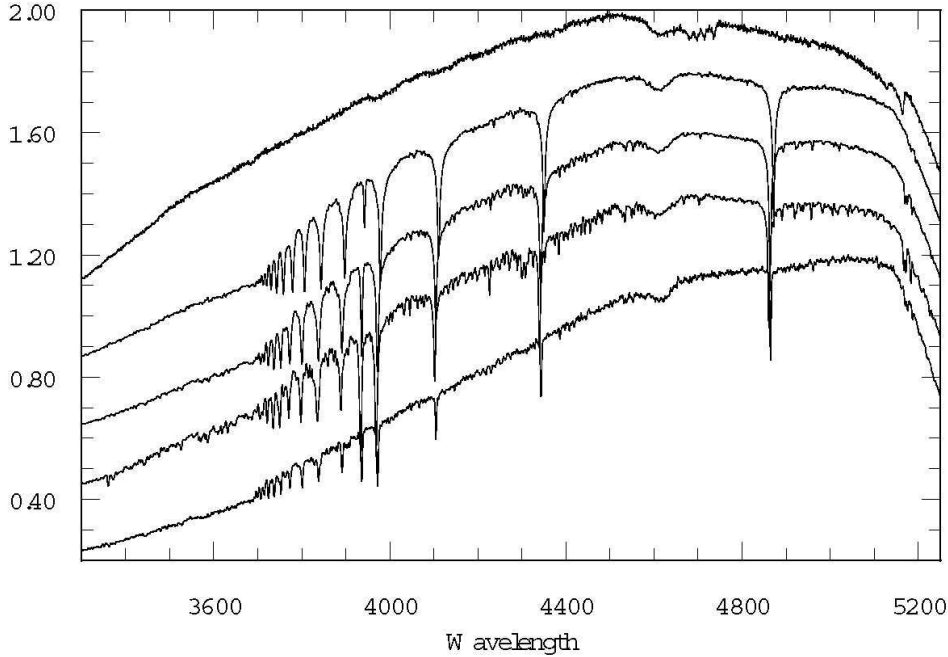,angle=0,clip=,height=9cm}}}
\caption[]{Raw extracted blue 2.3m spectra. See text for details}
\label{fig:f2}
\end{figure}

Fig.~\ref{fig:f2} shows examples of raw extracted blue spectra obtained with 
the Double Beam Spectrograph (DBS) on the ANU 2.3m telescope. The data has not 
been flat fielded. From the top, are the white dwarf LHS43 and the halo stars 
G64-12 (sdF0; 6500K/4.0/-3.4), 
HD 84937 (sdF5; 6200K/4.0/-2.3), G24-3 (sdF8; 5900K/4.2/-1.7), 
CD -38 245 (4800K/1.5/-4.1). The bracketed quantities are the approximate 
stellar parameters of the stars. 
The spectra are offset by 0.2 divisions for clarity.

\begin{figure}[h]
\centerline{\hbox{\psfig{figure=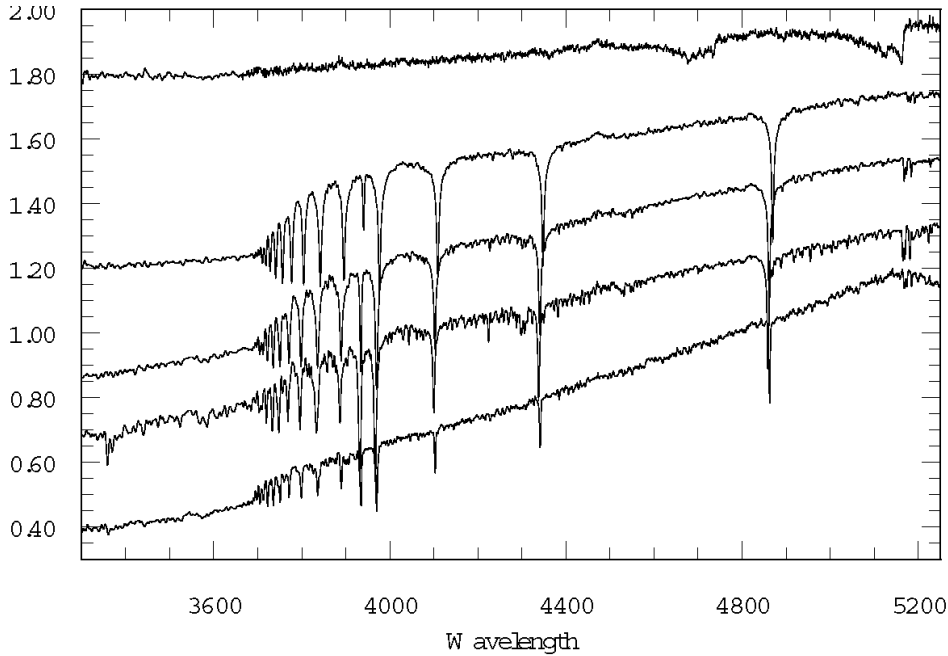,angle=0,clip=,height=9cm}}}
\caption[]{Same stars as in Fig.2 but divided by EG131.}
\label{fig:f3}
\end{figure}

Fig.~\ref{fig:f3} shows the same raw spectra divided by the spectrum of EG131.
The weak carbon bands in LHS43 and the Balmer jumps in the halo stars
are much more obvious and easier to measure in the divided spectra.

\section{Theoretical fluxes- model synthetic spectra}

Line-blanketed fluxes are available for the ATLAS9 and MARCS grid of model 
atmospheres. The \citet{mun2005} ATLAS9 synthetic spectra at 1\AA\/ resolution
are particularly useful. The MARCS sampled photospheric fluxes, available
on their website (http://marcs.astro.uu.se/), are provided with a warning 
that they are not likely to give a good representation of the integrated fluxes 
in a limited wavelength region. But they look adequate for the lowest 
metallicities.  

\begin{figure}[h]
\centerline{\hbox{\psfig{figure=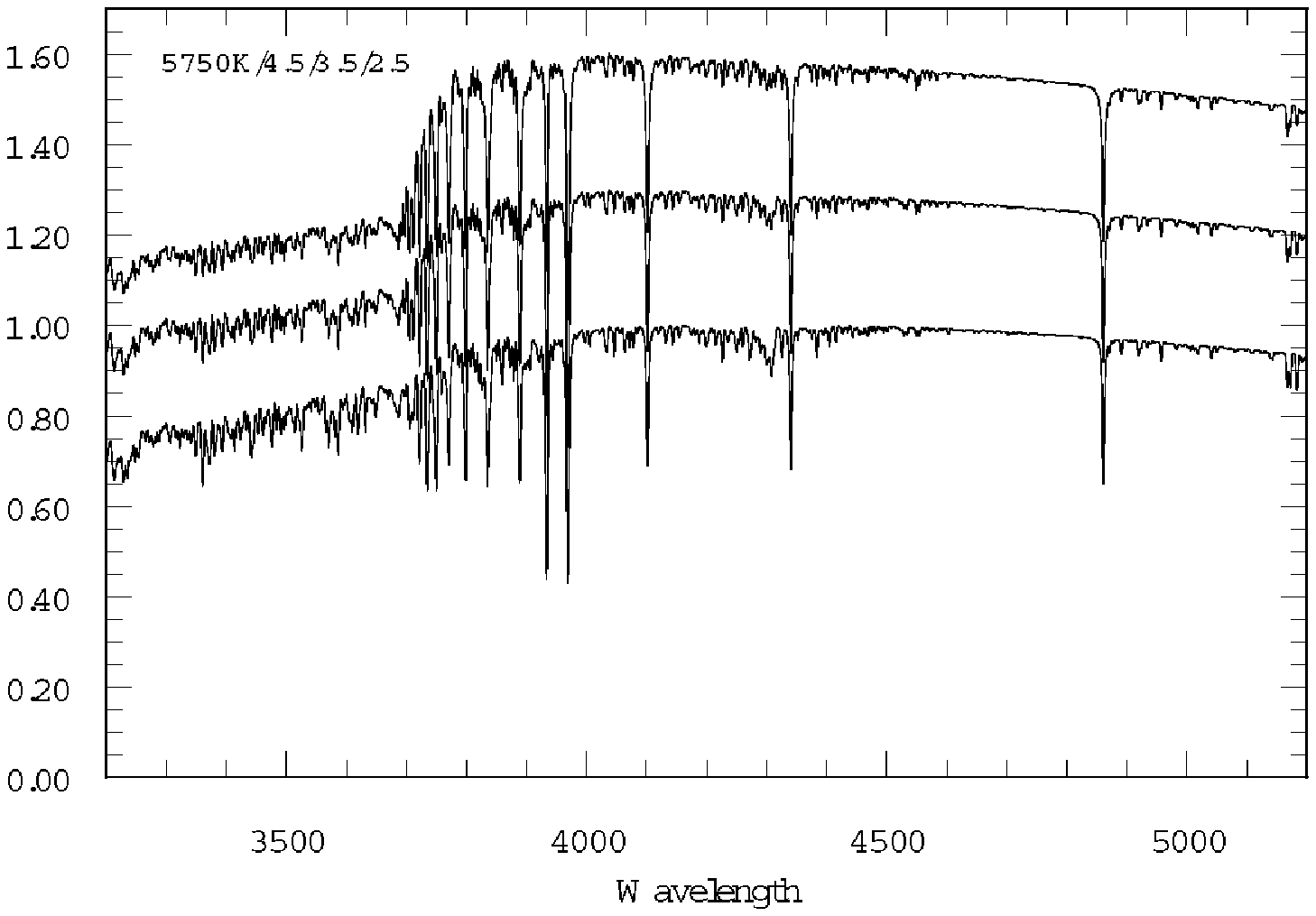,angle=0,clip=,height=9cm}}}
\caption[]{Synthetic flambda spectra for [Fe/H]=-2.5, 5750K and 
log g=4.5, 3.5, 2.5}
\label{fig:f4}
\end{figure}

Fig.~\ref{fig:f4} shows Munari synthetic flambda spectra for an abundance of
[Fe/H] = -2.5, temperature 5750K and three different effective gravities.
The spectra have been smoothed to 3\AA\/ resolution the same as the DBS data.
The Balmer discontinuity at 3636\AA\/ can be seen, but the slope of the continuum 
redward of this limit makes it difficult to measure it.

\begin{figure}[h]
\centerline{\hbox{\psfig{figure=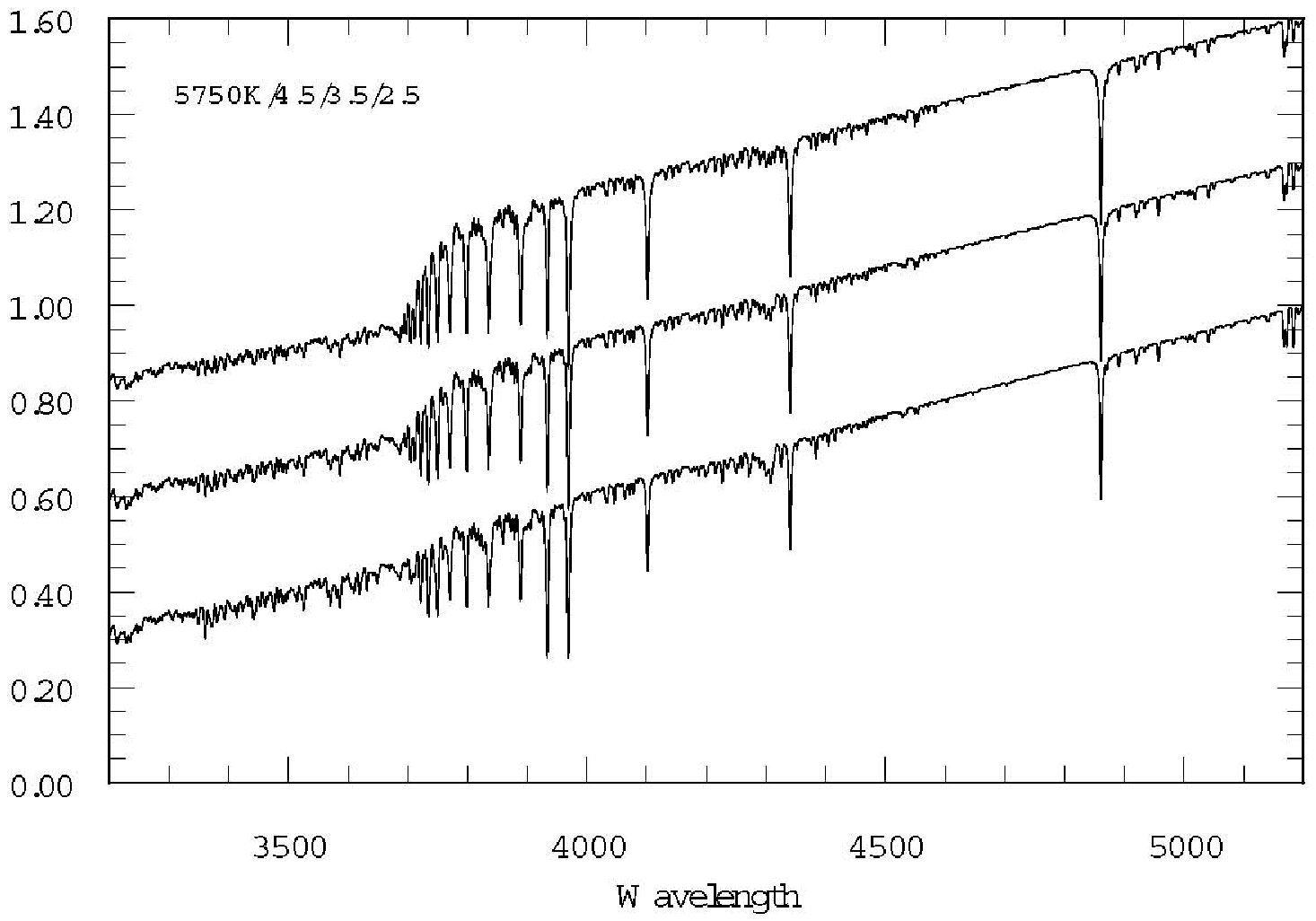,angle=0,clip=,height=9cm}}}
\caption[]{11800K BB divided spectra for [Fe/H]=-2.5, 5750K and log 
g=4.5, 3.5, 2}
\label{fig:f5}
\end{figure}

\begin{figure}[h]
\centerline{\hbox{\psfig{figure=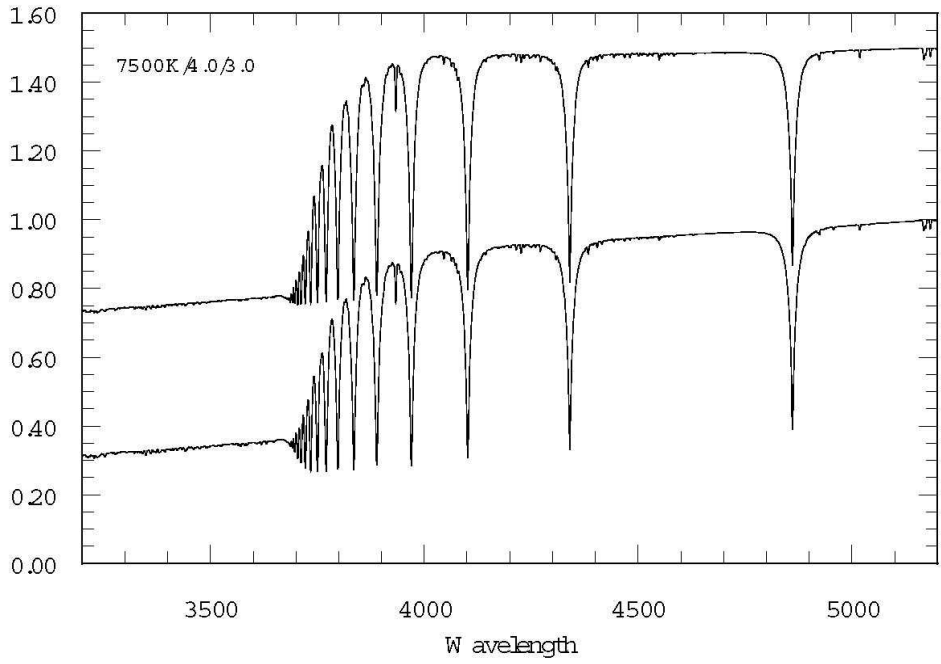,angle=0,clip=,height=9cm}}}
\caption[]{11800K BB divided spectra for [Fe/H]=-2.5, 7500K and 
log g=4.0, 3.0.}
\label{fig:f6}
\end{figure}

\begin{figure}[h]
\centerline{\hbox{\psfig{figure=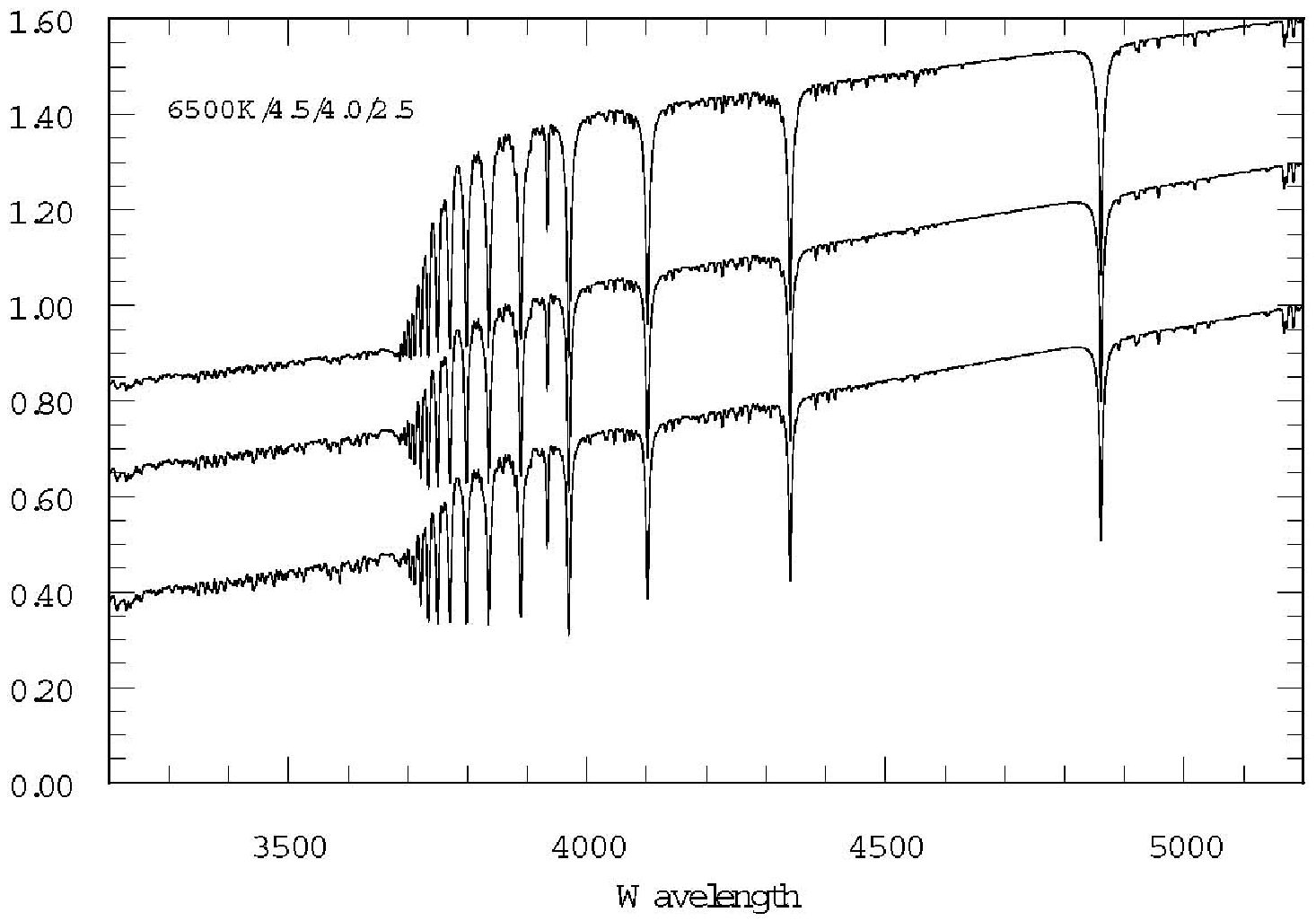,angle=0,clip=,height=9cm}}}
\caption[]{11800K BB divided spectra for [Fe/H]=-2.5, 6500K and 
log g=4.5, 4.0, 2.5.}
\label{fig:f7}
\end{figure}

\begin{figure}[h]
\centerline{\hbox{\psfig{figure=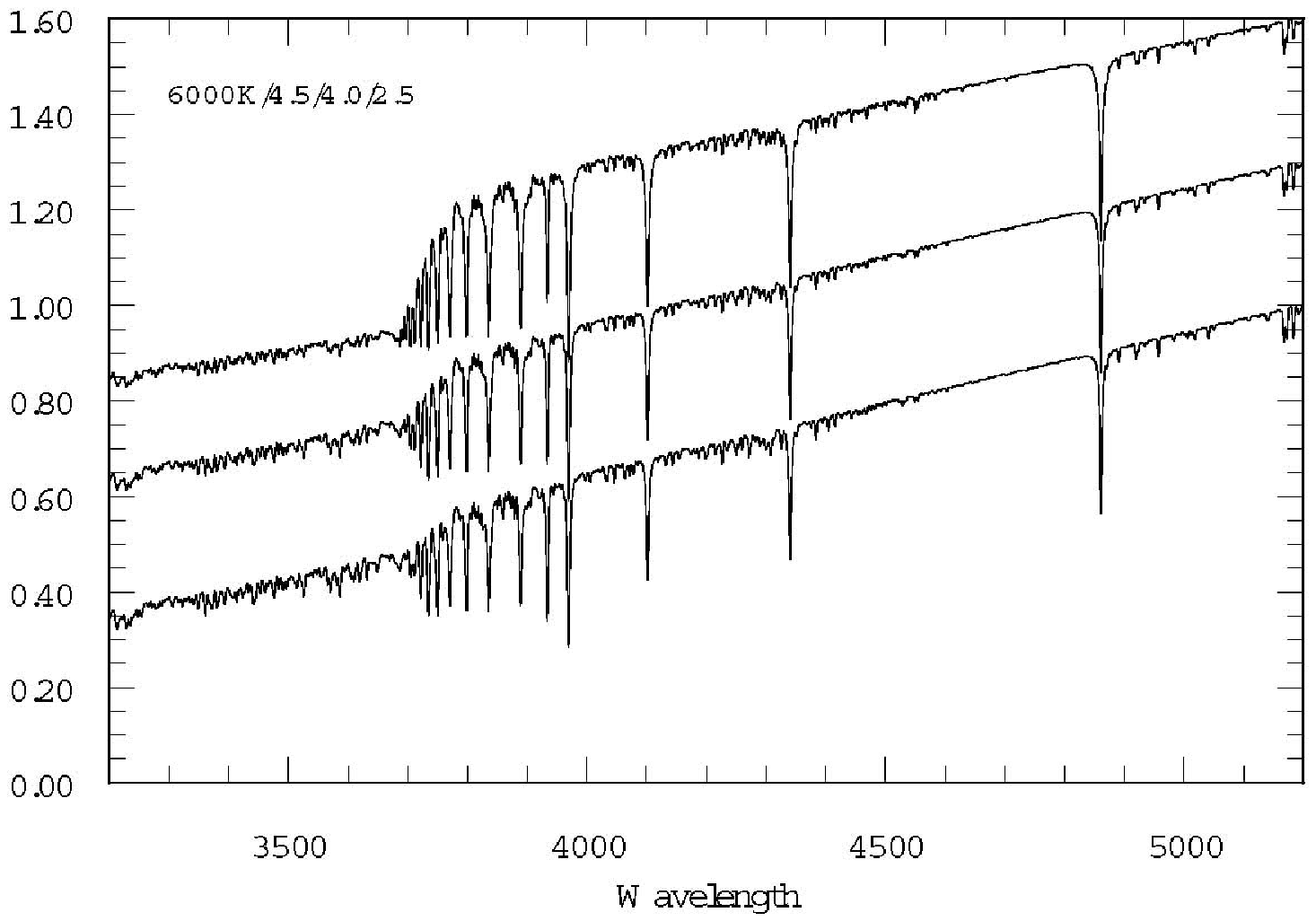,angle=0,clip=,height=9cm}}}
\caption[]{11800K BB divided spectra for [Fe/H]=-2.5, 6000K and 
log g=4.5, 4.0, 2.5.}
\label{fig:f8}
\end{figure}

\begin{figure}[h]
\centerline{\hbox{\psfig{figure=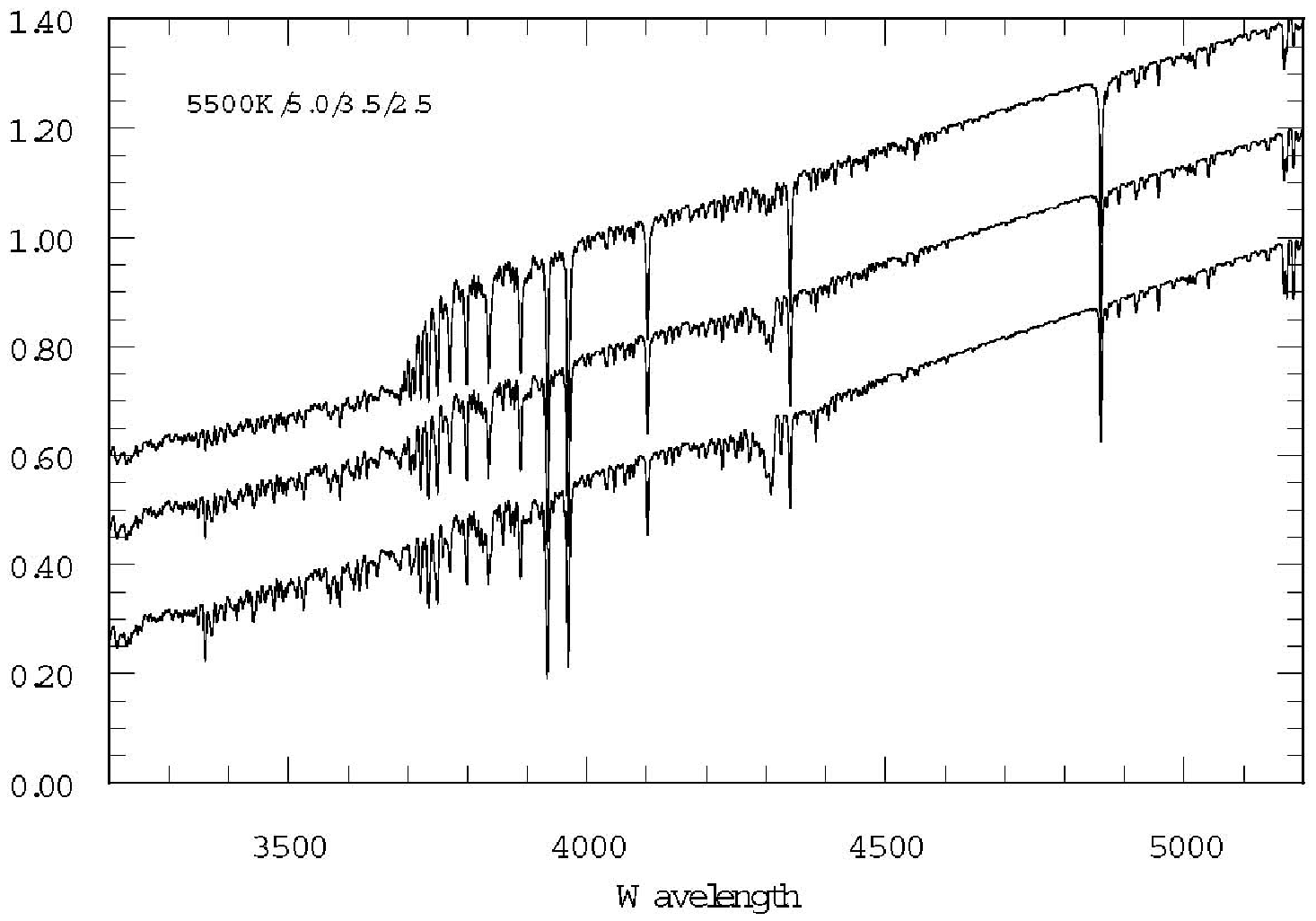,angle=0,clip=,height=9cm}}}
\caption[]{11800K BB divided spectra for [Fe/H]=-2.5, 5500K and 
log g=5.0, 3.5, 2.0.}
\label{fig:f9}
\end{figure}

\begin{figure}[h]
\centerline{\hbox{\psfig{figure=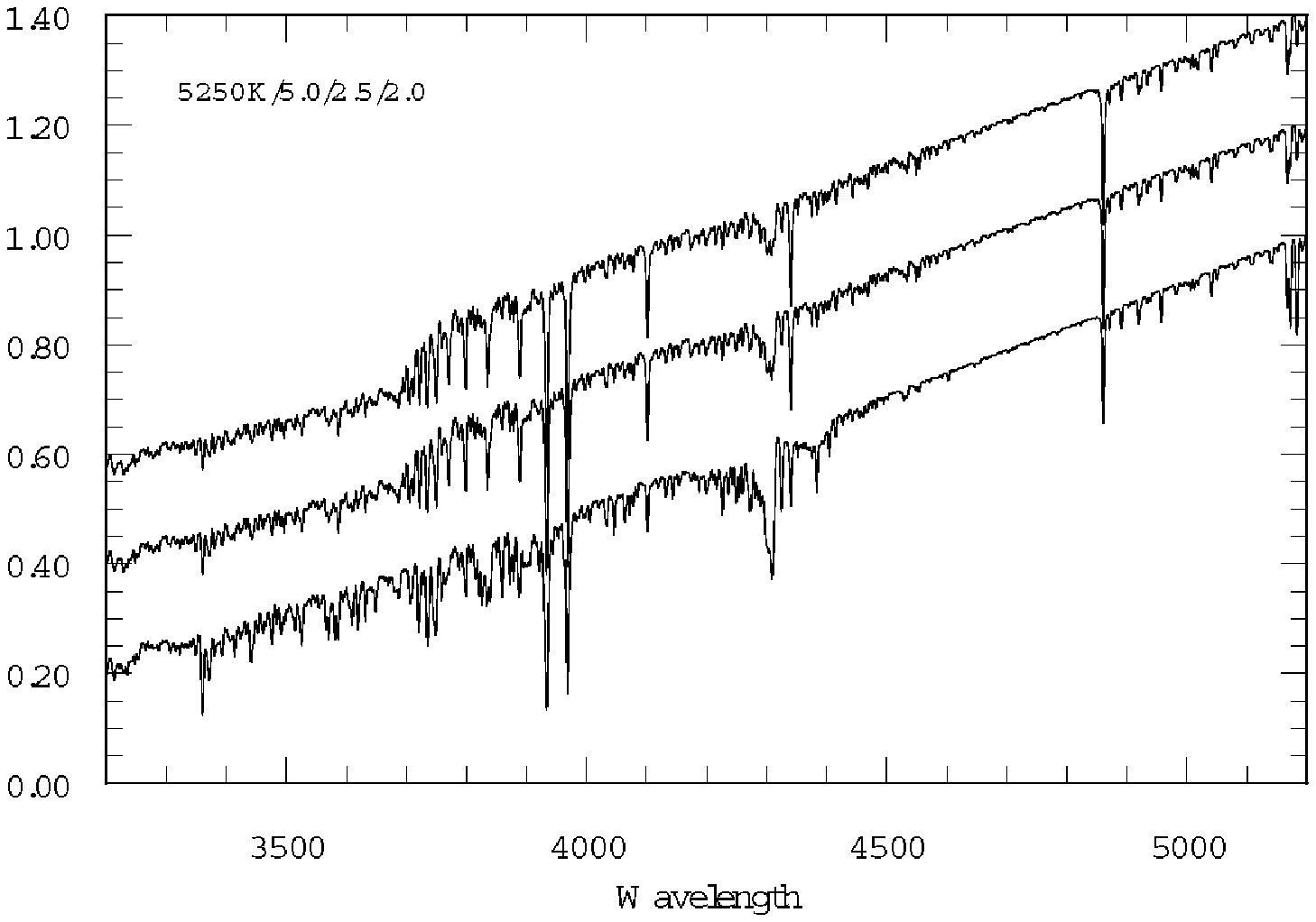,angle=0,clip=,height=9cm}}}
\caption[]{11800K BB divided spectra for [Fe/H]=-2.5, 5250K and 
log g=5.0, 2.5, 2.0.}
\label{fig:f10}
\end{figure}

The synthetic spectra can also divided by a 11800K blackbody to measure 
the theoretical Balmer jumps more reliably.
Fig.~\ref{fig:f5} are the same 5750K spectra divided by a 11800K black-body. 
Fig.~\ref{fig:f6} to ~\ref{fig:f10} show similarly divided Munari synthetic 
spectra for several more temperatures and sets of gravities appropriate for 
main-sequence, giant-branch and horizontal or asymptotic giant branch stars. 
For each temperature plot, the divided spectra have been normalized at 
5300\AA\/ and are plotted with equal offsets for better visibility. 
One can draw straight lines through the continua on the 
various figures to see how linear the continua of the divided observed and 
synthetic spectra are and how well defined the Balmer jumps are, 
even when they are small. 

\section{Balmer jump measurements}
\subsection{Theoretical spectra}
The model spectra were wavelength scrunched and then smoothed to approximate 
the observed medium resolution spectra. The Munari spectra were smoothed to 
5\AA\/ resolution, the MARCS fluxes to 6\AA\/ resolution. All model spectra were
divided by a 11800 black-body, plotted up
and the Balmer jumps measured by hand as for the observed spectra.
The size of the Balmer jump in magnitudes are given in Table 1 and Table 2.  
The depth of the H$\delta$ line was also measured as it is a good temperature
indicator, but for comparison with observations it is necessary to carefully
match the resolution whereas the Balmer jump measurement is not as sensitive
to the resolution. The theoretical b-y, B-V, V-R and V-I colors were also 
computed for the synthetic spectra. 

\small
\begin{deluxetable}{rrrrrrrrrrr}
\tablecolumns{11}
\tablewidth{0pc}
\tablecaption{MUNARI et al spectra}
\tablehead{
\colhead{} & \colhead{} &  \multicolumn{3}{c}{[Fe/H]=-1} & \multicolumn{3}{c}{[Fe/H]=-1.5} &  \multicolumn{3}{c}{[Fe/H]=-2.5} \\
\cline{3-11} 
\colhead{Te} & \colhead{log g}& \colhead{BJ}& \colhead{b-y} & \colhead{V-I} & \colhead{BJ} &\colhead{b-y} &\colhead{V-I}& \colhead{BJ}& \colhead{b-y} &\colhead{V-I}}
\startdata
6500 & 4.5 & 0.40 & 0.289 & 0.545 & 0.36 & 0.293 & 0.551 & 0.31 & 0.300 & 0.556 \\
6500 & 4.0 & 0.52 & 0.279 & 0.532 & 0.48 & 0.283 & 0.538 & 0.44 & 0.290 & 0.543 \\
6500 & 3.5 & 0.67 & 0.269 & 0.519 & 0.63 & 0.272 & 0.523 & 0.59 & 0.278 & 0.528 \\
6250 & 4.5 & 0.30 & 0.319 & 0.600 & 0.28 & 0.323 & 0.605 & 0.23 & 0.330 & 0.611 \\
6250 & 4.0 & 0.41 & 0.311 & 0.591 & 0.39 & 0.314 & 0.597 & 0.34 & 0.321 & 0.602 \\
6250 & 3.5 & 0.52 & 0.303 & 0.580 & 0.51 & 0.306 & 0.585 & 0.46 & 0.311 & 0.590 \\
6000 & 4.5 & 0.23 & 0.349 & 0.656 & 0.18 & 0.354 & 0.662 & 0.17 & 0.362 & 0.668 \\
6000 & 4.0 & 0.31 & 0.344 & 0.650 & 0.28 & 0.347 & 0.656 & 0.24 & 0.354 & 0.661 \\
6000 & 3.5 & 0.42 & 0.338 & 0.643 & 0.39 & 0.340 & 0.647 & 0.33 & 0.346 & 0.652 \\
5750 & 4.5 & 0.16 & 0.381 & 0.716 & 0.13 & 0.387 & 0.722 & 0.09 & 0.397 & 0.729 \\
5750 & 4.0 & 0.23 & 0.377 & 0.712 & 0.20 & 0.381 & 0.717 & 0.14 & 0.389 & 0.722 \\
5750 & 3.5 & 0.31 & 0.373 & 0.706 & 0.28 & 0.375 & 0.711 & 0.22 & 0.382 & 0.716 \\
5500 & 4.5 & 0.11 & 0.413 & 0.778 & 0.09 & 0.422 & 0.786 & 0.03 & 0.437 & 0.796 \\
5500 & 4.0 & 0.16 & 0.412 & 0.776 & 0.13 & 0.417 & 0.781 & 0.07 & 0.427 & 0.788 \\
5500 & 3.5 & 0.23 & 0.410 & 0.772 & 0.18 & 0.414 & 0.778 & 0.12 & 0.420 & 0.783 \\
5250 & 4.5 & 0.07 & 0.447 & 0.848 & 0.07 & 0.459 & 0.856 & 0.00 & 0.482 & 0.872 \\
5250 & 3.0 & 0.20 & 0.450 & 0.842 & 0.16 & 0.453 & 0.848 & 0.11 & 0.458 & 0.853 \\
\enddata 
\end{deluxetable} 

\small
\begin{deluxetable}{rrrrrrrrrrr}
\tablecolumns{11}
\tablewidth{0pc}
\tablecaption{MARCS statistical line opacity spectra}
\tablehead{
\colhead{} & \colhead{} &  \multicolumn{3}{c}{[Fe/H]=-1} & \multicolumn{3}{c}{[Fe/H]=-1.5} &  \multicolumn{3}{c}{[Fe/H]=-2.5} \\
\cline{3-11} 
\colhead{Te} & \colhead{log g}& \colhead{BJ}& \colhead{b-y} & \colhead{V-I} & \colhead{BJ} &\colhead{b-y} &\colhead{V-I}& \colhead{BJ}& \colhead{b-y} &\colhead{V-I}}
\startdata
6500 & 4.5 & 0.40 & 0.285 & 0.540 & 0.39 & 0.290 & 0.546 & 0.33 & 0.297 & 0.553 \\
6500 & 4.0 & 0.55 & 0.277 & 0.528 & 0.50 & 0.281 & 0.535 & 0.47 & 0.287 & 0.541 \\
6500 & 3.5 & 0.67 & 0.267 & 0.515 & 0.64 & 0.270 & 0.520 & 0.61 & 0.277 & 0.525 \\
6250 & 4.5 & 0.33 & 0.313 & 0.595 & 0.30 & 0.318 & 0.601 & 0.26 & 0.326 & 0.607 \\
6250 & 4.0 & 0.43 & 0.307 & 0.587 & 0.40 & 0.312 & 0.593 & 0.34 & 0.318 & 0.599 \\
6250 & 3.5 & 0.54 & 0.300 & 0.576 & 0.51 & 0.303 & 0.582 & 0.47 & 0.309 & 0.587 \\
6000 & 4.5 & 0.30 & 0.342 & 0.651 & 0.22 & 0.349 & 0.657 & 0.16 & 0.357 & 0.664 \\
6000 & 4.0 & 0.36 & 0.339 & 0.646 & 0.30 & 0.343 & 0.652 & 0.24 & 0.350 & 0.657 \\
6000 & 3.5 & 0.42 & 0.334 & 0.639 & 0.39 & 0.337 & 0.644 & 0.34 & 0.344 & 0.649 \\
5750 & 4.5 & 0.24 & 0.371 & 0.711 & 0.19 & 0.379 & 0.716 & 0.10 & 0.392 & 0.724 \\
5750 & 4.0 & 0.28 & 0.371 & 0.706 & 0.23 & 0.376 & 0.712 & 0.15 & 0.385 & 0.718 \\
5750 & 3.5 & 0.34 & 0.368 & 0.703 & 0.30 & 0.372 & 0.708 & 0.24 & 0.378 & 0.713 \\
5500 & 4.5 & 0.24 & 0.403 & 0.776 & 0.15 & 0.412 & 0.780 & 0.06 & 0.429 & 0.789 \\
5500 & 4.0 & 0.27 & 0.403 & 0.772 & 0.15 & 0.409 & 0.777 & 0.09 & 0.422 & 0.784 \\
5500 & 3.5 & 0.30 & 0.403 & 0.770 & 0.20 & 0.408 & 0.774 & 0.13 & 0.416 & 0.780 \\
5250 & 4.5 & 0.22 & 0.437 & 0.850 & 0.11 & 0.445 & 0.852 & 0.00 & 0.472 & 0.864 \\
5250 & 3.5 & 0.22 & 0.441 & 0.842 & 0.09 & 0.444 & 0.845 & 0.06 & 0.458 & 0.853 \\
\enddata     	        	   
\end{deluxetable}

\begin{figure}[h]
\centerline{\hbox{\psfig{figure=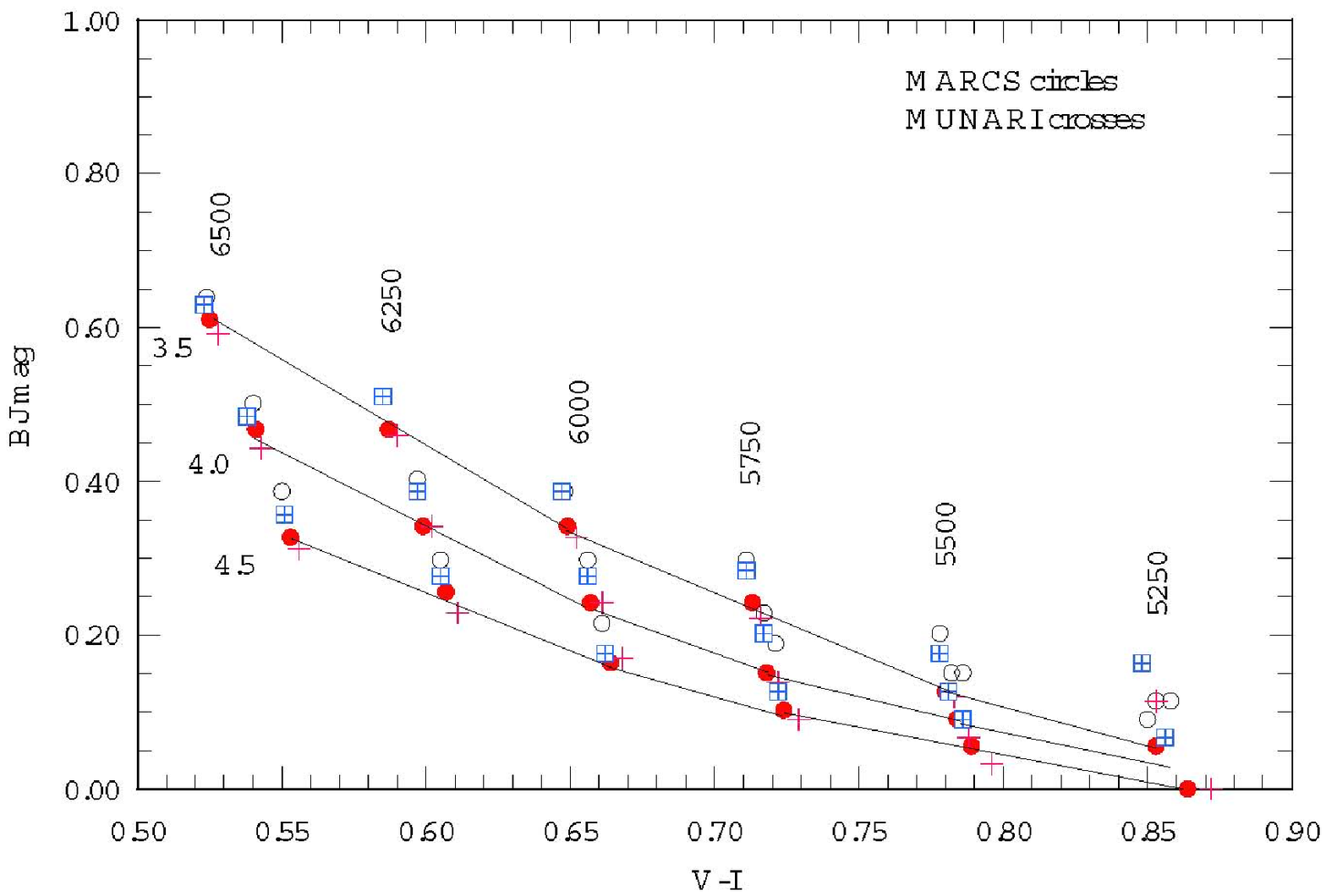,angle=0,clip=,height=9cm}}}
\caption[]{Measured Balmer jumps versus V-I color for model spectra for
[Fe/H]=-1.5 and -2.5. Connecting lines are drawn through points for the 
same gravity for the [Fe/H]=-2.5 spectra.}
\label{fig:f11}
\end{figure}

In Fig. 11 are plotted the measured Balmer jumps for the Munari spectra 
([Fe/H]=-2.5, cross; [Fe/H]=-1.5 crossed open square) and
the MARCS spectra ([Fe/H]=-2.5, filled circle; [Fe/H]=-1.5 open circle).
There is good general agreement for the two grids, except for the coolest
and strongest line spectra where as noted above, the MARCS sampled photospheric 
fluxes are not likely to give a good representation of the integrated flux in a 
limited wavelength region. Full spectrum synthesis is needed for the MARCS models 
to properly compare them. However, both sets of theoretical spectra show that for 
temperatures above 5500K the Balmer jump can readily distinguish main sequence 
and subgiant branch stars, 
while for temperatures at around 5000K and 5250K, only gravities below log g=3.0 
can be distinquished from log g=4.5 main sequence stars. Cooler than 5000K
the Balmer jump is virtually impossible to measure except in the most extreme
metal-poor stars.  

The model spectra show that the measured Balmer jump increases
as the metallicity increases due probably to the greater strength of metal-line
blanketing in the UV than in the blue.

\subsection{Observations}
There are several excellent spectral libraries that can be used for
Balmer jump measurements and comparison with the models.  
The best library is probably the Next Generation Spectral Library of STIS data by
\citet{gregg05}(http://lifshitz.ucdavis.edu/\~mgregg/gregg/ngsl/download.html). 
Preliminary versions of the spectra have been kindly provided by Michael Gregg and 
Jesus Maiz Appelaniz. The \citet{miles06} library 
(http://www.ucm.ed/info/Astrof/miles/miles.html)
is of simlar quality and resolution but has shorter wavelength coverage (3500\AA\/
- 7500\AA\/. .

Table 3 lists the measured Balmer jumps of the STIS spectra together with 
the Hipparcos Mv and colors computed from the spectra.

\begin{deluxetable}{lrrrrrr}
\tablecolumns{7}
\tablewidth{0pc}
\tablecaption{balmer jumps measured from STIS spectra}
\tablehead{
\colhead{Star} & \colhead{[Fe/H]}& \colhead{Mv}& \colhead{BJ} & \colhead{b-y} & \colhead{V-I} & \colhead{V-R}}
\startdata
bd092860  &      &  4.31 & 0.39 & 0.413 & 0.798 & 0.393 \\ 
bd174708  & -1.7 &  4.09 & 0.39 & 0.316 & 0.629 & 0.306 \\ 
bd292091  & -2.1 &  5.38 & 0.16 & 0.372 & 0.731 & 0.352 \\ 
bd371458  & -1.6 &  2.73 & 0.20 & 0.409 & 0.774 & 0.373 \\ 
bd413931  & -1.8 &  6.05 & 0.08 & 0.415 & 0.809 & 0.401 \\ 
bd423607  & -2.0 &  5.51 & 0.19 & 0.344 & 0.639 & 0.308 \\ 
bd511696  & -1.4 &  5.58 & 0.19 & 0.394 & 0.778 & 0.380 \\ 
bd592723  & -1.7 &  5.22 & 0.33 & 0.337 & 0.708 & 0.340 \\ 
bd720094  & -1.6 &  3.99 & 0.40 & 0.303 & 0.671 & 0.315 \\ 
bd-122669 & -1.5 &  4.04 & 0.65 & 0.227 & 0.500 & 0.236 \\ 
cd-3018140& -2.2 &  4.26 & 0.42 & 0.298 & 0.570 & 0.270 \\ 
g115-58   & -1.4 &  3.62 & 0.37 & 0.325 & 0.696 & 0.335 \\ 
g029-023  & -2.0 &  2.78 & 0.39 & 0.338 & 0.684 & 0.334 \\ 
g88-27    & -1.9 & -0.82 & 0.36 & 0.316 & 0.654 & 0.312 \\ 
hd002665  & -2.0 &  4.34 & 0.11 & 0.537 & 1.005 & 0.494 \\ 
hd002857  & -1.6 &  1.24 & 1.38 & 0.085 & 0.216 & 0.089 \\ 
hd006755  & -1.6 &  2.16 & 0.15 & 0.489 & 0.957 & 0.467 \\ 
hd016031  & -1.8 &  4.46 & 0.37 & 0.305 & 0.629 & 0.298 \\ 
hd019445  & -2.0 &  5.10 & 0.27 & 0.329 & 0.681 & 0.319 \\ 
hd044007  & -1.7 &  1.62 & 0.09 & 0.548 & 1.045 & 0.511 \\ 
hd045282  & -1.5 &  2.33 & 0.17 & 0.448 & 0.861 & 0.422 \\ 
hd063791  & -1.7 & -0.89 & 0.00 & 0.588 & 1.127 & 0.549 \\
hd087140  & -1.6 &  2.18 & 0.21 & 0.461 & 0.946 & 0.452 \\ 
hd094028  & -1.3 &  4.63 & 0.30 & 0.327 & 0.685 & 0.324 \\ 
hd111721  & -1.1 &  0.56 & 0.13 & 0.507 & 0.981 & 0.480 \\ 
hd128279  & -2.1 &  1.90 & 0.18 & 0.449 & 0.908 & 0.432 \\ 
hd132475  & -1.3 &  3.73 & 0.28 & 0.363 & 0.778 & 0.365 \\ 
hd134439  & -1.5 &  6.74 & 0.04 & 0.424 & 0.857 & 0.429 \\ 
hd134440  & -1.5 &  7.08 & 0.02 & 0.516 & 1.059 & 0.538 \\ 
hd163346  &      &  0.30 & 0.99 & 0.380 & 0.839 & 0.403 \\ 
hd163810  & -1.4 &  5.00 & 0.20 & 0.384 & 0.745 & 0.368 \\ 
hd165195  & -2.2 &  4.02 &      & 0.894 & 1.533 & 0.774 \\   
hd184266  & -1.8 &  0.17 & 0.55 & 0.418 & 0.817 & 0.400 \\ 
hd284248  & -1.4 &  4.77 & 0.36 & 0.321 & 0.659 & 0.316 \\ 
hd345957  & -1.3 &  3.75 & 0.33 & 0.345 & 0.742 & 0.347 \\ 
\enddata
\end{deluxetable}

Table 4 lists the measured Balmer jumps of the MILES spectra together with the 
atmospheric parameters from \citet{miles07} and the b-y and B-V colors computed 
from the spectra.    

\small
\begin{deluxetable}{rrrrrrr}
\tablecolumns{7}
\tablewidth{0pc}
\tablecaption{Balmer jumps measured from MILES spectra}
\tablehead{
\colhead{Star} & \colhead{[Fe/H]}& \colhead{Te}& \colhead{log g} & \colhead{BJ}& \colhead{b-y}& \colhead{B-V}}
\startdata
hd002665 & -2.0 & 5013 & 2.35 & 0.22 & 0.509 & 0.720 \\
hd002796 & -2.3 & 4950 & 1.36 & 0.28 & 0.520 & 0.724 \\
hd019445 & -2.1 & 5920 & 4.40 & 0.26 & 0.345 & 0.445 \\
hd045282 & -1.4 & 5350 & 3.20 & 0.22 & 0.458 & 0.658 \\
hd046703 & -1.6 & 6000 & 0.40 & 1.20 & 0.302 & 0.421 \\
hd064090 & -1.8 & 5450 & 4.45 & 0.08 & 0.424 & 0.608 \\ 
hd074000 & -2.0 & 6170 & 4.20 & 0.37 & 0.322 & 0.407 \\ 
hd084937 & -2.2 & 6230 & 4.00 & 0.45 & 0.301 & 0.374 \\
hd085773 & -2.2 & 4460 & 1.00 & 0.00 & 0.737 & 1.046 \\
hd088609 & -2.6 & 4513 & 1.30 &      & 0.676 & 0.941 \\
hd094028 & -1.5 & 5950 & 4.20 & 0.31 & 0.347 & 0.462 \\
hd103095 & -1.4 & 5030 & 4.60 & 0.09 & 0.487 & 0.754 \\
hd122563 & -2.6 & 4570 & 1.12 & 0.09 & 0.653 & 0.903 \\
hd140283 & -2.5 & 5690 & 3.60 & 0.30 & 0.361 & 0.474 \\
hd165195 & -2.2 & 4470 & 1.10 &      & 0.830 & 1.219 \\
hd187111 & -1.8 & 4260 & 0.60 &      & 0.809 & 1.215 \\
hd188510 & -1.6 & 5490 & 4.70 & 0.23 & 0.403 & 0.591 \\
hd218502 & -1.8 & 6030 & 3.80 & 0.43 & 0.309 & 0.414 \\
hd218857 & -1.9 & 5080 & 2.40 & 0.18 & 0.492 & 0.705 \\
hd219617 & -1.4 & 5880 & 4.00 & 0.26 & 0.348 & 0.478 \\
hd221170 & -2.1 & 4470 & 1.00 &      & 0.712 & 1.049 \\
hd237846 & -2.6 & 4960 & 1.80 & 0.20 & 0.489 & 0.668 \\
hd251611 & -1.7 & 5350 & 3.80 & 0.15 & 0.474 & 0.664 \\ 
hd284248 & -1.6 & 6025 & 4.20 & 0.39 & 0.328 & 0.422 \\ 
\enddata
\end{deluxetable}

In Fig. 12, the Mv versus b-y diagram for the STIS spectra is shown together 
with a halo isochrone. Some of the Mv values are poorly determined because of
large uncertainties in the parallax and there are at least 4 stars whose 
absolute magnitudes are in disagreement with their Balmer jump measurements
shown in Fig.13. 

\begin{figure}[h]
\centerline{\hbox{\psfig{figure=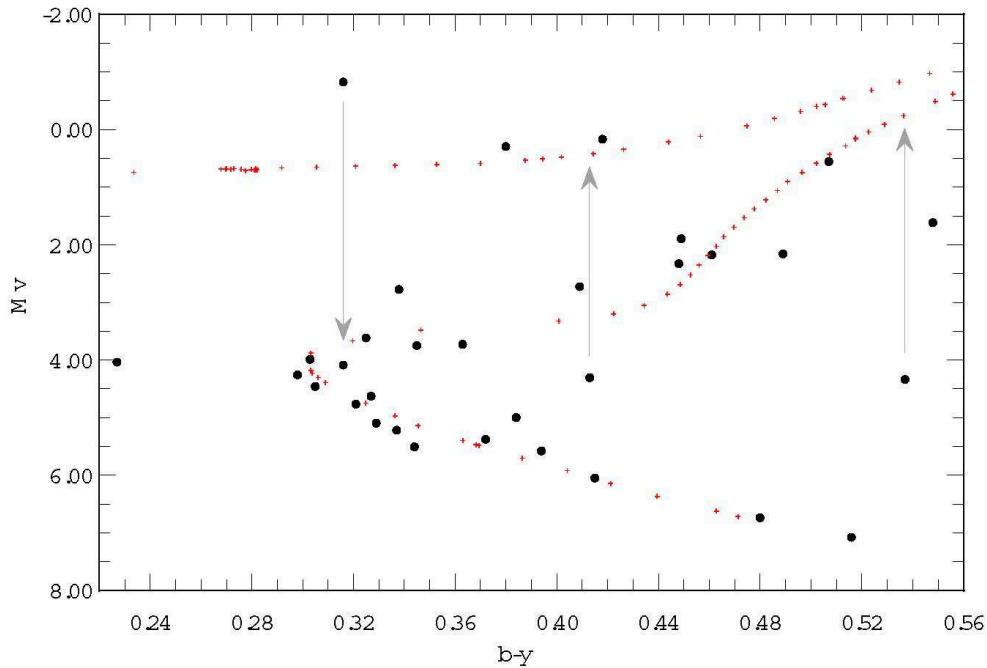,angle=0,clip=,height=9cm}}}
\caption[]{The Mv versus b-y diagram for the selection of STIS spectra of 
low metallicity stars listed in Table 3. A halo isochrone is shown for orientation.
The lines with arrow heads indicate the appropriate shifts for those stars whose 
Mv values are inconsistent with their measured Balmer jumps.}
\label{fig:f12}
\end{figure}

\begin{figure}[h]
\centerline{\hbox{\psfig{figure=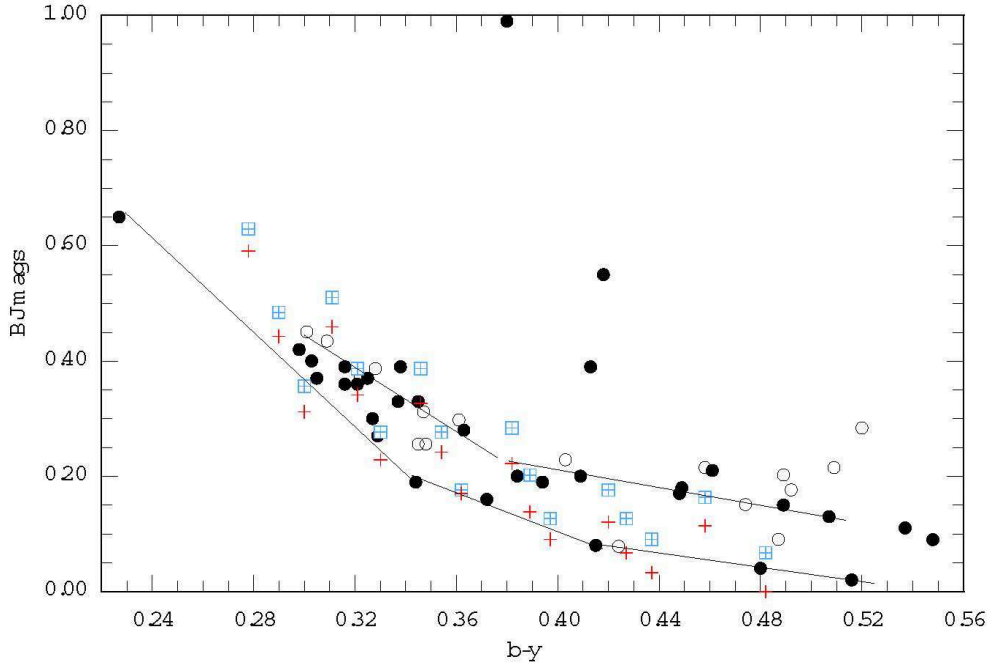,angle=0,clip=,height=9cm}}}
\caption[]{Measured Balmer jumps versus b-y color for the STIS stars 
(closed circles) and MILES spectra (open circles). Lines have 
been drawn to connect the stars on the main-sequence, the subgiant and the 
giant branch as identified in Fig 12. The same relation for the Munari spectra for 
[Fe/H]=-1.5 plotted in Fig. 11 is also shown. }.   
\label{fig:f13}
\end{figure}

In Fig 13 are plotted the observed Balmer jumps for the STIS sample versus
b-y. Lines have been drawn through the main-sequence stars, the subgiant and 
giant branch stars as indicated from Fig 12. The stars whose Mv values differ
significantly from their Balmer jumps are marked in Fig 12. The theoretical 
values for the Munari [Fe/H]=-1.5 and -2.5 spectra are also shown for 
log g = 3.5, 4.0 and 4.5 and temperatures T${_e}$=6500K, 6250K, 6000K, 
5750K, 5500K and 5250K. The Balmer jumps from the synthetic spectra are in 
good agreement with the observed loci of metal-poor stars. Precise derivation 
of the gravities of individual stars involves knowledge of the metallicity
and temperature (and the reddening if fluxes are used), which is beyond 
the scope of this paper.

\section{Summary}
Dividing extracted raw spectra by the spectrum of a warm near-blackbody object
such as EG131, L745-46a, LHS43 or LHS4043 results in spectra whose continua 
above and below the Balmer discontinuity are close to straight lines thus
making it easy to measure the size of the Balmer jump accurately. 

The recommendation is to observe a warm near-black-body star at least once
a run for each grating setting used then divide all extracted raw spectra by 
a template made from the normalised extracted blackbody spectrum by
removing any obvious lines or bands it may have or fitting the continuum. 
In order to accurately measure the theoretical Balmer jumps from synthetic
spectra such as those of \citet{mun2005}, it is also important to divide 
the synthetic spectra by the normalised theoretical blackbody spectrum of 
similar black-body temperature to the template star used. 

Having followed these recommendations and measured the Balmer jump, 
the effective gravity can be derived from the model spectra. 
If the effective temperature of the FGK star is known to within 100K 
the Balmer jump will yield an effective gravity to about 0.2 dex for 
the hotter stars and 0.5 dex for the cooler stars, sufficient 
precision to determine whether the star is a main-sequence, 
giant branch or horizontal/asymptotic giant branch star.

Best spectrophotometric results are achieved if an atmospheric dispersion 
corrector is used or if the spectrograph is rotated to put the parallactic 
angle along the slit. This is not essential to measure the monochromatic 
Balmer jump but it does ensure that the spectra can be accurately calibrated 
onto a relative absolute flux scale and that other temperature sensitive 
colors can be derived.





\section{Acknowledgments}
I am grateful to Dr Lajos Balazs, Director of Konkoly Observatory and 
Dr Katalin Olah for their hospitality during the writing of this paper. 
I am also grateful to Dr Fiorella Castelli for providing the synthetic 
spectra; to Dr Santi Cassisi for computing additional isochrones and to
Dr Michael Gregg and Dr Jesus Maiz Appelaniz for the preliminary NGSL STIS spectra. 



{\it Facilities:} \facility{SSO: 2.3m (DBS)} \facility{HST: STIS}

\end{document}